\documentclass[showkeys,nofootinbib,
 jor,
 amsmath,amssymb,
]{revtex4-2}
\usepackage{graphicx}% Include figure files
\usepackage{dcolumn}% Align table columns on decimal point
\usepackage{bm}% bold math
\usepackage{xcolor}
\begin{document}

%\preprint{AIP/123-QED}

\title{Observable imprints of primordial gravitational waves on the temperature anisotropies of the Cosmic Microwave Background} 
%\footnote{Error!}}% Force line breaks with \\
%\thanks{Footnote to title of article.}
\author{Miguel-Angel Sanchis-Lozano and Veronica Sanz}
 \email{miguel.angel.sanchis@ific.uv.es, veronica.sanz@uv.es}
\affiliation{Departamento de Física Teórica and Instituto de Física Corpuscular, CSIC-University of Valencia%\\This line break forced with \textbackslash\textbackslash
}


\begin{abstract}
We examine the contribution of tensor modes, in addition to the dominant scalar ones, on the temperature anisotropies of the cosmic microwave background (CMB). To this end, we analyze in detail the temperature two-point angular correlation function $C(\theta)$ from the  {\it Planck} 2018 dataset, focusing on large angles ($\theta \gtrsim 120^{\circ}$) corresponding to small $\ell$ multipoles. A hierarchical set of infrared cutoffs are naturally introduced to the scalar and tensor power spectra of the CMB
by invoking an extra Kaluza-Klein dimension compactifying
at about the GUT scale between the Planck epoch and the start 
of inflation. We associate this set of lower scalar and tensor cutoffs with
the parity of the multipole expansion of the $C(\theta)$ function. By fitting the {\it Planck} 2018 data we compute the
the multipole coefficients thereby reproducing the well-known odd-parity preference in angular correlations seen 
by all three satellite missions COBE, WMAP and {\it Planck}.
Our fits improve significantly once tensor modes are included in the analysis, hence providing a hint of the imprints of primordial gravitational waves on the temperature correlations observed in the CMB today. To conclude we suggest a relationship between, on the one hand, the lack of (positive) large-angle correlations and
the odd-parity dominance in the CMB and, on the other hand, the effect of primordial gravitational waves on the CMB temperature anisotropies.
\end{abstract}


\maketitle

\section{\label{sec:level1}Introduction}

In an inflationary scenario, primordial energy density inhomogeneities, due to the unavoidable quantum fluctuations of the primordial field, are the seed of the primary temperature CMB anisotropies, as well as the large scale structure of our observable universe today. Usually, temperature anisotropies 
are mainly attributed to the scalar modes of the inflaton field, while tensor modes are expected to be subdominant, therefore hardly distinguishable from the former. On the other hand, common wisdom usually states that polarization 
of the CMB remains as the main hope to detect primordial gravitational waves (PGW) produced in the very early universe~ \cite{Caprini:2018mtu} as only tensor -and not scalar - modes can produce B-modes of polarization~\cite{Chiocchetta:2020ylv}. Note however that disentangling B-modes of PGW from other 
non-cosmological sources (e.g. gravitational lensing) renders this method rather cumbersome so far. Nevertheless, future very-high precision measurements of the CMB polarization \cite{Bouchet(2011),Hanany(2019),Errard(2016)}
could uncover a PGW background in this way. 

In this work, we include the expected small but maybe still observable effect of the tensor modes on the CMB temperature correlations especially at large angles (low multipoles), in spite of the big uncertainties including the cosmic variance \cite{Copi(2010)}. Use will be made of a parity statistic \cite{Aluri(2012),Panda(2021)} to highlight the long-standing apparent odd-parity preference shown by data of all three satellite missions COBE \cite{Hinshaw(1996)} , WMAP \cite{Bennett(2003)} and {\it Planck} \cite{Planck(2018),Planck(2019)} in the multipole analysis of the two-point correlation function. Notice that such a parity imbalance, even to a small degree, questions the large-scale isotropy of the observable universe stemming from the cosmological principle. 

In fact, the standard cosmological model can be viewed as a phenomenological effective theory of an unknown underlying more general theory yet to be discovered. 
Discrepancies, anomalies or puzzles, which are emerging from observations with respect to a standard cosmology scenario \cite{Abdalla(2022),DiValentino(2021),Shaikh(2019),Muir(2018),Rassat(2014)}
may have a systematic origin, or can be due to statistical fluctuations. Their persistence, however, along different probes implying uncorrelated errors, strongly suggest the need for new physics beyond the minimal standard model in cosmology and elementary particle physics.

\subsection{Angular correlations of the CMB and the Sachs-Wolfe effect}

Two categories of temperature fluctuations observed in the CMB can be distinguished according to
the cosmic time evolution: (a) primary anisotropies, prior to decoupling, and (b) secondary anisotropies, 
developing as photons propagate from the surface of the last scattering to the observer (us). The former
include temperature fluctuations due to photon propagation under metric perturbations originated by inhomogeneities of the matter field at the time of recombination, the so-called
Sachs-Wolfe (SW) effect. This effect shows up at rather large angles from directions in the celestial sphere, while secondary anisotropies (as well as Baryonic Acoustic Oscillations) rather affect quite small angles and have been successfully accounted for by standard cosmology.

The SW effect will play an important role in our study to account for the temperature angular distribution, showing a lack of (positive) correlations at large angle (i.e. $\theta \gtrsim 60^{\circ}$) together with an apparent parity imbalance. 
Usually, under some simplifying assumptions, a plateau is expected from this effect in the angular power spectrum at low $\ell$ ($\lesssim 30$), while a sawtooth shape favoring odd-$\ell$ peaks is actually observed,later interpreted in this paper.
%as the core of our results.

\subsection{Parity imbalance seen in the CMB}

As is well known, Nature is parity violating, e.g. in the electroweak sector of the Standard Model where only left-handed fermions are active \cite{Thomson(2013)}. In this context, it is natural to ask whether Nature would again violate parity through some gravitational processes, and if this feature could shed light on the very early universe itself. In fact, a variety of sources of gravitational parity violation have been considered in the literature , see e.g.\cite{Lue(1998),Alexander(2009)}.
%, from fundamental quantum gravity effects  
%to rolling inflationary axions, or pseudoscalar inflation. 
Any of these could have have left an imprint on the net helicity of the gravity wave background, namely the preferred excitation of one circular polarization over another. 

In the present study we have followed a different path, started in a previous paper (see Ref.\cite{Sanchis(2022b)}) where a certain degree
of parity breaking showing up in correlations at large angle of the CMB radiation was envisaged. To this end, fundamental fermionic fields (making up a composite scalar inflaton) were introduced satisfying periodic and antiperiodic conditions
on a Hubble radius in real space. In this way, two distinct infrared cutoffs, associated with integer and half-integer
Fourier modes of the fluctuating field, emerged as a source of parity breaking
in angular correlations. In particular, a preference for odd multipoles at large angle comes out naturally in angular two-point correlations (see appendix). 

In this paper we will not resort to the existence of fermionic fields as fundamental components of a composite scalar inflaton   
to provide the required periodic and antiperiodic conditions, as done in Ref.\cite{Sanchis(2022b)}. Rather, we shall assume that 
in the very early universe a pre-inflationary scalar field satisfies certain boundary conditions on an extra spatial dimension {\it \`{a} la Kaluza-Klein} (just one for simplicity), to be adressed in more detail in section IV. Thereby, distinct comoving scales for infrared cutoffs (commonly related to the inverse radius of an extra dimension compactified as a circle ) come into play in temperature angular correlations for both scalar and tensor modes.

In this regard, the power spectrum itself, defined 
by the Fourier transform of the primordial fluctuation spectrum, is often parameterized as
\begin{equation}\label{eq:PS}
    P^S(k)= A^S\ \biggl[\frac{k}{k_{\ast}}\biggr]^{n_s-1}
\end{equation}
where $n_s$ is the scalar spectral index and $k_{\ast}$ the pivot scale. The above spectrum, referred exclusively to scalar modes, would be perfectly scale free (i.e. $n_s = 1$) if the Hubble parameter $H_{\rm inf}$ were
strictly constant during inflation. If $H_{\rm inf}$ evolves
slowly, a slight deviation of the spectral index from one is expected and indeed the observations show that that $n_s = 0.9649 \pm 0.0042$ \cite{ParticleDataGroup:2022pth}.

A near scale-free $P^S(k)$ would have been generated as modes with comoving wavenumbers $k$ successively crossed the
Hubble radius and classicalized, later reentering into the Hubble horizon of the observable Universe once inflation ended. Usually no lower limit is assumed in the $P^S(k)$ spectrum so that in numerical computations of observables the integration range 
over Fourier modes is taken between zero and infinity.

On the other hand, a tensor power spectrum $P^T(k)$ is usually
parametrized as
\begin{equation}\label{eq:PT}
    P^T(k)= A^T\ \biggl[\frac{k}{k_{\ast}}\biggr]^{n_T}
\end{equation}
where the tensor spectral index $n_T$ is expected to be small but not vanishing. 

From Eqs.(\ref{eq:PS}) and (\ref{eq:PT}), the  
tensor-to-scalar ratio $r$ is defined as
\begin{equation}\label{eq:r}
    r= \frac{A^T}{A^S}= \frac{P^T(k_{\ast})}{P^S(k_{\ast})}
\end{equation}
Current limits on $r$ severely constrain many models of inflation
and we will later check that the tensor contributions computed
in this work comply with such conditions.

\section{two-point angular correlation function of the CMB}

All three COBE, WMAP and  {\it Planck} satellite missions have observed that the temperature angular distribution of the CMB is remarkably homogeneous across the sky, with anisotropies of order 1 part in $10^5$. This observation is in fact one of the main arguments in favor of an inflationary scenario 
in cosmology to solve the so-called horizon problem, 
together with the flatness and monopole problems \cite{Kolb(1994)}.

A powerful test of these
fluctuations relies on the two-point angular correlation function $C(\theta)$~\footnote{See also Refs.~\cite{Gangui(1993),Kamionkowski(2010)} for a discussion on higher-order correlation functions.}, defined as the ensemble product
of the temperature differences with respect to the average temperature $T_0$ from all pairs of directions
in the sky defined by unitary vectors $\vec{n}_1$ and $\vec{n}_2$:
\begin{equation}\label{eq:CTT}
C(\theta)=\langle \frac{\delta T(\vec{n}_1)}{T_0}\frac{\delta T(\vec{n}_2)}{T_0} \rangle\;,
\end{equation}
where $\theta \in [0,\pi]$ is the angle defined by the scalar product $\vec{n}_1 \cdot \vec{n}_2$.

The information contained in the angular power spectrum of the CMB
is basically the same as in the correlation function, but the
latter highlights the behaviour at large angles (small $\ell$) where a sizable disagreement between theoretical expectations and observations has been found. In this work we
focus on the analysis of angular correlations searching specifically for 
imprints from the very early universe on the temperature fluctuations.

The temperature two-point correlation function is usually expanded as
\begin{equation}\label{eq:C2}
C(\theta)=\ \sum_{\ell \ge 2} \frac{(2\ell+1)}{4\pi}\ C_{\ell}\ P_{\ell}(\theta)
\end{equation}
where $P_{\ell}(\theta)$ is the order-$\ell$ Legendre polynomial, and the sum extends from $\ell=2$ since the monopole and dipole contributions have been removed from the analysis.

In the following, we will ignore the transfer function and consider only the SW effect as the main source 
of primary anisotropies, as expected  
on scales larger than $\simeq 1^{\circ}$. Hence, the multipole coefficients of Eq.(\ref{eq:C2}) can be computed 
in the limit of a flat power spectrum $P^S(k)$, as
\begin{equation}\label{eq:Cellnocutoff}
C_{\ell}^S\ =\ \frac{2N^S}{\pi}\ \int_0^{\infty}\ du\ \frac{j_{\ell}^2(u)}{u}\ =\ \frac{N^S}{\pi\ell(\ell+1)}
\end{equation}
where $j_{\ell}$ is the spherical $\ell$-Bessel function and $N^S$ stands for a normalization factor (related to the amplitude $A^S$ in Eq.(\ref{eq:PS})) to be determined from the fit to observational data.

An overall agreement between the behaviour
of $C(\theta)$ and observational points can be achieved
if the $C_{\ell}$ coefficients comply with the SW plateau condition ($C_{\ell} \sim 1/\ell(\ell+1)$ at small $\ell$). However, the $\chi^2_{\rm d.o.f.}$ resulting from the fit
turns out to be quite unsatisfactory as, e.g., 
positive correlations arise at large angle, in contrast to
the observed lack of large-angle correlations for $\theta \gtrsim 60^{\circ}-70^{\circ}$ \cite{Abdalla(2022)} 
among other anomalies \cite{Zhao(2015)}.

\section{Single infrared cutoff in the scalar power spectra}

In order to improve the above-mentioned unsatisfactory fit, an infrared cutoff $k_{\min}$ was introduced {\it ad hoc} to the CMB power scalar spectrum in Ref.~\cite{Melia(2018)}, implying a  lower limit $u_{\min}$ in the integral of the multipole
$C_{\ell}$ coefficients in Equation (\ref{eq:Cellnocutoff}). 
\begin{equation}\label{eq:Cellcutoff}
C_{\ell}^S\ =\ \frac{2N^S}{\pi}\ \int_{u_{\rm min}}^\infty\ du\ \frac{j_{\ell}^2(u)}{u}\;,
\end{equation}
where $k_{\rm min}=u_{\rm min}/r_d$ and $r_d$ denotes the co-moving distance to the last scattering surface.
The authors of ref.~\cite{Melia(2018)} 
introduced the infrared cutoff essentially in
a heuristic way whose purpose was removing the unseen
(positive) correlations at large angle expected 
in standard cosmology. Later, a 
theoretical interpretation of $k_{\rm min}$ was given 
as the first oscillation mode to leave the Planck domain \cite{Melia(2019)}
within a linearly expanding (without inflation) universe (see \cite{Melia(2012)}).

In Ref.~\cite{Melia:2021tvx}, this study was consistently extended focusing on the low-$\ell$ region of the power spectrum itself,  providing an infrared cutoff value compatible with~\cite{Melia(2018)}.
Furthermore, in Ref.\cite{Sanchis(2022)} a suggestive connection between the lack 
of correlation at large angles provided by $k_{\rm min}$ 
and the odd-parity preference was shown. Indeed, the observed downward tail of the $C(\theta)$ function 
at large angles ($\theta \gtrsim 120^{\circ}$) was nicely
reproduced, while keeping the good 
behaviour of $C(\theta)$ over the whole examined angular range, $4^{\circ}< \theta \leq 180^{\circ}$.

Notice, however, that  
a heuristic tuning of the multipole coefficients ($\ell \lesssim 10$) was required in Ref.\cite{Sanchis(2022)} in order to
reach a good agreement with data. In a following work  \cite{Sanchis(2022b)}, {\em two} lower cutoffs (instead of one) were introduced in the scalar power spectrum, providing the desired odd-parity preference of $C(\theta)$
but avoiding so many fit parameters.

Indeed, based on causality arguments two maximum correlation lengths ($\lambda_{\rm max}^{\rm even}$ and $\lambda_{\rm max}^{\rm odd}$ affecting
even and odd parity multipoles respectively) were
put forward in Ref.\cite{Sanchis(2022b)} in correspondence with
two different boundary conditions. Then two comoving wavenumbers ($k_{\rm min}^{\rm even}$ and $k_{\rm min}^{\rm odd}$) were defined accordingly for odd and even parity modes.
The existence of such a doublet of (periodic and antiperiodic) boundary conditions was attributed to fundamental fermionic fields
making up a composite inflaton \cite{Samart(2022)}.

Following essentially this idea, however no fermionic fields will be invoked in the present paper to get two different energy scales for
the cutoffs. Rather, we will rely on a Kaluza-Klein model with an extra spatial dimension, 
so that appropriate boundary conditions of
a scalar field (at the very early universe) would yield even and odd fields (regarding their Fourier expansion in $k$ modes) in the observable four-dimensional universe. 
Moreover, a similar pattern will be assumed for tensor fluctuations 
contributing to the angular correlation function, to be incorporated to out analysis as a second step.

But first, let us examine a simple extra dimension model 
providing a theoretical framework and motivation for the phenomenological
assumptions made so far, to be further developed in the subsequent sections.

\section{Model set-up in Extra-Dimensions}

In this section we provide a specific scenario where those cutoffs would arise naturally. We will build a model in five-dimensions (5D) of space-time, which would lead to a 4D low energy theory when the fifth dimension is compactified. This compactification would lead to the appearance of 4D Kaluza-Klein modes whose spectrum would dictate the cutoffs we mentioned.  

%\subsection{A five-dimensional realisation}~\label{sec:hall}

Let us assume that Nature at some high scale is five-dimensional (5D). For simplicity, let us also assume for the moment that there is no curvature and those 5D are flat. This situation would be described by a 5D Minkowski metric
\begin{equation}\label{eq:metric}
    ds^2 = dt^2-dx_i dx^i - dz^2 ,
\end{equation}
where $i$=1, 2 and 3, and we denote with $z$ the 5-th dimension.

The Universe appears to be 4D, though, and very precise tabletop experiments do confirm that the gravitational attraction between two objects of masses $m_1$ and $m_2$ is the one expected from Newton's gravity in 4D, namely
$V(r) \propto m_1 m_2/r$. 

%Indeed

However, if gravity acted over a larger number of spacetime dimensions ($d=4+n$), the law would be modified to
$$V(r) \propto \frac{m_1 m_2}{r^{1+n}} .$$ 
which is not ruled out provided that the size of the extra-dimension is smaller than a few tenths of $\mu$m, see Ref.~\cite{tabletop} for a recent and most precise limit. 

%This is ruled out by everyday observations, which nevertheless do not exclude the existence of a small-size extra-dimension, as long as its size is smaller than a few tenths of $\mu$m, see Ref.~\cite{tabletop} for a recent and most precise limit. 

Tiny extra-dimensions can arise through a process called compactification. There are various options to realize the compactification mechanism, including the presence of fluxes in Calabi-Yau manifolds. Depending on the configuration of the geometry and the fluxes, one can end up with 
different possibilities for a compact extra dimension.

For simplicity, let us assume that the extra-dimension is simply a segment of size $L$.
%an circle of radius $R$. 
Fields propagating in this geometry could be factorized in the so-called Kaluza-Klein decomposition. For any  field, one can write their expression as follows
$$\Phi(x^\mu, z) = \phi(x^\mu) \, f(z),$$
where $x^\mu$, with $\mu=0\ldots 3$, are the 4D coordinates and $z$ is the fith dimension, $z\in [0,L]$.  

For a scalar/fermion field in extra-dimensions, their 4D fields $\phi(x^\mu)$ would satisfy the Klein-Gordon/Dirac equation of motion and the fifth component, $f(z)$, would satisfy a wavefunction equation which depends on the geometry and boundary conditions at the two extremes of the interval.  For example, the field $\Phi$ could have Neumann or Dirichlet boundary conditions, namely:
\begin{eqnarray}
\partial_z \Phi|_{z=z_0} &=&0 \textrm{   (Neumann, or $+$)}\\ \nonumber \Phi (z_0) &=& 0 \textrm{ (Dirichlet, or $-$).}
\end{eqnarray}

Those boundary conditions can be imposed in both extremes of the interval, at $z_0=0$ and $z_0=L$, leading to different options for fields in the extra-dimension. From the 4D standpoint, these fields would appear as an infinite tower of 4D fields (Kaluza-Klein tower) with masses determined by the boundary conditions. 

Fields with  Neumann boundary conditions at both ends ($z=0$, $z=L$) would be called $(+,+)$ and would exhibit a massless zero mode. Fields with other boundary conditions, $(\pm,\mp)$ and $(-,-)$, would have a KK spectrum with a lowest KK state ($n=0$) with a non-zero mass, providing an IR cutoff in their spectrum. Those four options for boundary conditions at $z=0$ and $L$ have different properties under the parity $z\to -z$. Indeed,
\begin{equation}
    (+,+) \textrm{ and } (-,-) \textrm{ are {\it even},  and }  (+,-) \textrm{ and } (-,+) \textrm{ are {\it odd}. }
\end{equation}

In flat geometries like the one described by the metric in Eq.(\ref{eq:metric}), the mass spectrum of fields with boundary conditions $(+,-)$ and $(-,-)$ would be related as follows,$\frac{m^{(-,-)}_n}{m^{(+,-)}_n}=\frac{2n+2}{2n+1}$, where $n=0,1\ldots$. For these 5D fields, the lowest mass in their spectrum ($n=0$) provides a natural IR cutoff for their physical behaviour. The IR cutoffs could be represented by a letter $k_S$  and related by a simple factor 2,
$$\frac{k_S^{\rm even}}{k_S^{\rm odd}}=2,$$
precisely the factor we would consider when evaluating their impact on the scalar two-point correlation in the CMB. Here we have labelled the field with $(-,-)$ as even and $(+,-)$ as odd, following the convention in the next sections. We have also added the subscript $S$ to indicate the field we consider is scalar.
 
Besides scalar or fermion fields, spin-two fields are unavoidably present in any spacetime geometry. From the 4D standpoint, the Kaluza-Klein tower of the graviton spin-two field must contain  a massless state, responsible for 4D gravity, and a tower of massive fields (KK-gravitons) with the same quantum numbers as the massless graviton. 

The condition of a massless state implies that the tensor field satisfies $(+,+)$ boundary conditions. This choice determines the rest of the graviton spectrum. The massive spin-two fields would follow a 4D Fierz-Pauli lagrangian (describing a massive spin-two state) and their interactions with other species would be driven by the coupling to the stress tensor~\cite{Randy}. 

This construction can be generalized to metrics with curvature in the extra-dimension. Geometries with curvature in the fifth dimension have been employed in various applications, including the AdS/CFT correspondence~\cite{Maldacena:1997re}, or in holographic approaches for QCD~\cite{Hirn:2005nr} and for electroweak interactions~\cite{Hirn:2006nt,Hirn:2006wg}. In those more generic cases, analytic expressions of the KK spectrum cannot be obtained but in Ref.~\cite{Hirn:2007bb} closed expressions for the overall IR behaviour of the spectrum  (sum rules) were derived. Also, let us mention that successful inflation could be achieved in these kinds of scenarios by assuming the inflaton is a pseudo-Goldstone boson and its potential is generated by the KK contributions of fields in the extra-dimension~\cite{Croon:2014dma}.

Moreover, one can relate the KK tower of spin-two fields, KK-gravitons, to other fields (fermions or scalars) propagating in the extra-dimension. In flat geometries, the first KK-graviton (coming from a $(+,+)$ field) would have its mass at the same value as the $(+,-)$ lowest mode of a scalar or fermion field, 
although this relation can be modified in the presence of curvature, typically leading to $k_{T}> k^{\rm odd}_{S}$,  as explained in Refs.~\cite{Randy,Dillon}.

The second KK-graviton state could also contribute to the tensor correlations. The mass would be double of the first KK graviton, leading to a relation for the first two states in the KK-graviton spectrum $k_T^{'} \simeq 2 k_T$.

To summarise, fields propagating in an extra-dimension  exhibit a 4D spectrum that depends on the geometry  and the boundary conditions. The impact of these fields on 4D observables can be computed as a sum over contributions of 4D Kaluza-Klein fields. Below the compactification scale, those contributions would be dominated by the lowest KK modes, whose masses provide  a natural IR cutoff. In the analysis of the CMB spectrum, we will consider that the cutoffs coming from the extra-dimensional scenario would be as follows
\begin{eqnarray} \label{eq:cutoffsXDIM}
    \textrm{Scalar cutoffs: }& & k_S^{\rm even} = 2 k_S^{\rm odd},  \nonumber\\
    \textrm{Tensor cutoffs (from KK gravitons): }& & k_T \gtrsim k_S^{\rm odd} \textrm{ and }  k_T^{'} = 2 k_T.
\end{eqnarray}

These (infrared) cutoffs will be taken into account in the next sections when computing correlation functions for the scalar and tensor perturbations.  
\vspace{2.cm}

\section{Double infrared cutoff in the scalar power spectrum}

Once given in the past section a theoretical background to the existence of infrared cutoffs in the power spectra, let us introduce the same notation as 
in Ref.\cite{Sanchis(2022b)} 
\begin{equation}\label{eq:u2}
k_{\rm min}^{\rm odd/even}=
\frac{u_{\rm min}^{\rm odd/even}}{r_d} 
\end{equation}
corresponding now to two
infrared cutoffs (instead of one) in the
scalar power spectrum.

Thus we rewrite the integral of Eq.(\ref{eq:u2}) as
\begin{equation}\label{eq:Cellcutoffs}
C_{\ell_{\rm odd/even}}^S\ =\ \frac{2N^S}{\pi}\ \int_{u_{\rm min}^{\rm odd/even}}^{\infty}\ du\ \frac{j_{\ell}^2(u)}{u}\;,
\end{equation}
where the lower limits of the integral are now 
$u_{\rm min}^{\rm odd/even}=k_{\rm min}^{\rm odd/even}/r_d$, 
thereby affecting differently the numerical values of
the odd and even coefficients (actually only at low $\ell$), therefore altering the shape of $C(\theta)$. 

From a best fit of $C(\theta)$ to the {\it Planck} 2018 data, the following values for the lower cutoffs were obtained in \cite{Sanchis(2022b)}: $u_{\rm min}^{\rm odd}= 2.67 \pm 0.31$ and $u_{\rm min}^{\rm even}=5.34 \pm 0.62$. In this work we will recompute these lower cutoffs but
incorporating tensor modes in our analysis, also imposing the relations from the compact extra-dimension in (\ref{eq:cutoffsXDIM}). The new numerical values for both $u_{\rm min}^{\rm even}(\rm scalar)$ will not differ greatly from the previous ones, but tensor fluctuations will certainly contribute to angular correlations, as we shall see soon.

Let us remark by now
that the improvement regarding the
parity dominance (affecting the downward tail in $C(\theta)$ at large angle) resulting from the introduction of two infrared cutoffs with respect to a single one, is rather limited because
the net effect is restricted to the first few multipoles: 
$\ell \lesssim u_{\rm min}$.
Let us mention in this regard the so-called ellipsoidal universe \cite{Cea(2022)}, where only the quadrupole term is actually modified, however yielding a noticeable effect in large-angle
correlations.

To overcome this drawback and extend further the
influence of the lower cutoff(s) on the multipole expansion of $C(\theta)$, we will next
include the contribution of ({\rm tensor}) modes thereby
modifying further the $C_{\ell}$, reaching higher $\ell$ values.

\section{Tensor modes and extra infrared cutoffs}

In this section we address the effect of primordial
gravity waves generated during inflation on the CMB temperature anisotropies due to tensor modes, which constitutes the main goal of this paper. To this end, and taking into account that we are
examining angular correlations, the following expression  
will be used to compute the tensor coefficient of the $\ell$ multipole according to Ref.\cite{Mukhanov(2005)} 
\begin{equation}\label{eq:Celltensor}
C_{\ell}^T=\ \frac{N^T\ (\ell-1)\ell(\ell+1)(\ell+2)}{2\pi}\ \int_0^{\infty}\ du\ \frac{j_{\ell}^2(u)}{u^5}=\ \frac{N^T}{15\pi}\ \frac{1}{(\ell+3)(\ell-2)},\ \ \ell > 2
\end{equation}
in an analogous way as $C_{\ell}^S$ in Eq.(\ref{eq:Cellnocutoff}), where the normalization factor $N^T$ 
is now related to the amplitude $A^T$ in Eq.(\ref{eq:PT}),  similarly as $N^S$ for the scalar modes.

Similarly to the scalar case and following the
arguments given in Section IV, we will assume that two lower
cutoffs equally apply to the above integral in Eq.(\ref{eq:Celltensor}) such that
\begin{equation}\label{eq:Celltensoroddeven}
C_{\ell\rm odd/even}^T\ =\ N^T \ 
\frac{(\ell-1)\ \ell\ (\ell+1)\ (\ell+2)}{2\pi} 
\int_{u_{\rm min}^{\rm odd/even}({\rm tensor})}^{\infty}\ 
du\ \frac{j_{\ell}^2(u)}{u^5}
\end{equation}
again distinguishing odd and even modes by different
lower cutoffs, satisfying the ratio  
$u_{\rm min}^{\rm even}({\rm tensor})=2u_{\rm min}^{\rm odd}({\rm tensor})$. 

%Let us remark again that any sizable effect on $C_{\ell}$ of (both scalar and tensor lower cutoffs is restricted to rather low multipoles,recovering the expected SW plateau-like behaviour at higher $\ell$, i.e. $\ell \gtrsim u_{\rm min}$.

%Moreover, the following relations between the scalar and tensor infrared cutoffs are assumed:
%\begin{eqnarray}\label{eq:STconditions}
%u_{\rm min}^{\rm odd}({\rm tensor}) &=&
%u_{\rm min}^{\rm even}({\rm scalar}) \\
%u_{\rm min}^{\rm even}({\rm tensor}) &=&
%u_{\rm min}^{\rm even}({\rm scalar})
%\end{eqnarray}

\vspace{1.7cm}

\section{Parity statistic analysis}

As commented in the Introduction, a possible connection between an $\lq\lq$odd
universe" (i.e. parity-breaking) and
the lack for large-angle correlations has been contemplated in the literature 
though without a clear
theoretical explanation yet Ref.\cite{Land(2005),Kim(2012),Schwarz(2016),Copi(2018)}. 
The deviation from even-odd parity balance 
in angular correlations can be studied by means of a parity statistic defined in Refs.~\cite{Aluri(2012),Panda(2021)}
\begin{equation}\label{eq:Qstat}
Q(\ell_{\rm max}^{\rm odd})=\frac{2}{\ell_{\rm max}^{\rm odd}-1}\ \sum_{\ell=3}^{\ell_{\rm max}^{\rm odd}}\ 
 \frac{D_{\ell-1}}{D_{\ell}}\ ;\ \ell_{max}^{\rm odd} \ge 3\;,
\end{equation}
where $\ell_{max}^{\rm odd}$ stands for the maximum odd multipole up to which the statistic is computed. 
Any deviation 
from unity of this statistic as a function of $\ell_{max}^{\rm odd}$ points to an even-odd parity imbalance: below unity implies odd-parity dominance and a downward tail at large-angle in the $C(\theta)$ plot, above one implies 
even-parity dominance and a upwards tail. As we will see soon, the parity statistics $Q(\ell_{\rm max}^{\rm odd})$ play a crucial role in our work 
assessing the different contributions of scalar and tensor modes to angular correlations.

\subsection{Physical scale of the infrared cutoffs}

From our previous fits to {\it Planck} 2018 data, the resulting scalar infrared cutoffs (as comoving wavenumbers) turned out to be of order:
\begin{equation}
    k_{\rm min} \simeq \frac{{\rm few}}{r_d} \approx {\rm few} \times 10^{-4}\ {\rm Mpc^{-1}}\ \to\  k_{\rm min} \simeq 
    {\rm few} \times 10^{-42}\ {\rm GeV}
\end{equation}
where the last figure is expressed in GeV units, 
which amounts to a very low value indeed. However, physical
cutoffs have to be obtained by dividing them by the scale factor, in particular at the time when compactification of the
extra dimension took place ($t_{\rm extra}$), likely at the very early universe. 
Therefore the corresponding physical wavenumber, basically set
by the inverse radius of the extra dimension, has to be computed 
as $k_{\rm min}/a(t_{\rm extra})$. 

Now assuming that the time
of compactification happened at some time between the end
of the Planck epoch $t_{\rm Planck}$ and the beginning of
inflation $t_{\rm init}$, the respective scale factors 
$a(t_{\rm Planck})\simeq 10^{-61}$ and $a(t_{\rm init})\simeq 10^{-56}$ given in \cite{Liu-Melia(2023)} lead to the
following physical cutoff range:
\begin{equation}\label{eq:kscale}
    \frac{k_{\rm min}}{a(t_{\rm extra})} \in 10^{19}-10^{14}\ {\rm GeV},\ \ t_{\rm Planck} < t_{\rm extra} < t_{\rm init}
\end{equation}
which interestingly contains the GUT scale.

Thus the existence of an infrared cutoff in the CMB temperature correlations is consistent with the assumption of an extra dimension with a high compactification scale.  
%and the resulting hierarchy of infrared cutoffs
%stemming from our analysis on , become thereby supported theoretically by invoking an extra dimension.  

\begin{figure*}[h]
\centering
\includegraphics[width=13.cm]{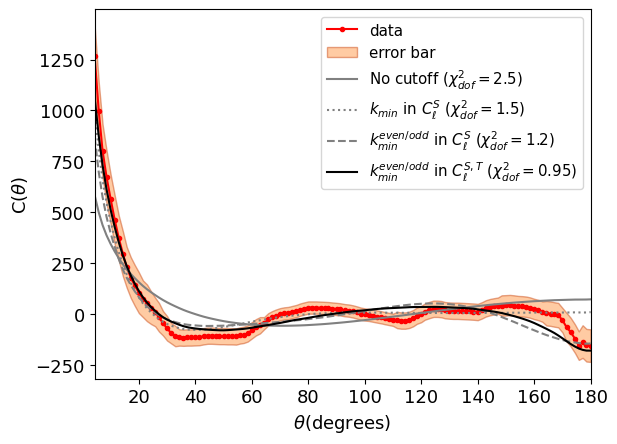}
\caption{\label{fig_1a} Temperature 2-point angular correlation function as a function of $\theta$ from
the fit to {\it Planck} 2018 data, for the four different assumptions discussed in the text. The shadowed area represents one $\sigma$ error bar.
}
\end{figure*}
\section{Final analysis and discussion}

Let us first summarize the successive steps made in this work
concerning the set of infrared cutoffs in the scalar and tensor power spectrum 
to improve the $\chi^2_{\rm d.o.f.}$ of the 
fits to {\em Planck} 2018 data points. 

\begin{itemize}\label{eq:items}
\item[a)] No infrared infrared cutoff is introduced to the
scalar power spectrum, and the multipole coefficients satisfy the SW-plateau, i.e., $C_{\ell}\sim 1/\ell(\ell+1)$, $\ell \lesssim 30$. 
\item[b)] A single infrared cutoff $k_{\rm min}$
 is introduced to the scalar power spectrum corresponding to $u_{\rm min}=4.5$ in the integral (\ref{eq:Cellcutoff}).
\item[c)] Two infrared cutoffs $k_{\rm min}^{\rm odd/even}$
 are introduced to the scalar power spectrum 
yielding two lower cutoffs
$u_{\rm min}^{\rm even}=2u_{\rm min}^{\rm odd} \simeq 5.4$ in
the integral (\ref{eq:Cellcutoffs}).
\item[d)] A set of infrared cutoffs $k_{\rm min}^{\rm odd/even}(\rm tensor)$ are further introduced to the tensor power spectrum, in addition to the scalar modes, according
to the pattern: $u_{\rm min}^{\rm even}({\rm tensor})=2u_{\rm min}^{\rm odd}({\rm tensor})$, as theoretically motivated in section IV. 
%Besides,the ratio $N^T/N^S$ plays a fundamental role in setting the relative contributions of scalar and tensor modes affecting 
%Eqs.(\ref{eq:Cellcutoffs}) and (\ref{eq:Celltensoroddeven}), respectively.

\end{itemize}

Figures 1 and 2 show the best $\chi^2_{\rm d.o.f.}$ fits of $C(\theta)$ and $Q({\ell}_{max})$ to {\it Planck} data. Let us remark that, under the conjecture that $u_{\rm min}^{\rm odd}({\rm tensor})=u_{\rm min}^{\rm even}({\rm scalar})$, 
incorporating the tensor contribution in our analysis does not mean increasing the number of free fit parameters, except for the extra normalization factor $N^T$, in Eq.(\ref{eq:Celltensoroddeven}).

From the above fits the value of the lower cutoff $u_{\rm min}^{\rm odd}({\rm scalar})=2.5$ is extracted, while the
values of all the other (both scalar and tensor) lower cutoffs become determined by the pattern associated with the compact extra dimension (i.e. multiplicative factors 2). Let us also note that this value of $u_{\rm min}^{\rm odd}({\rm scalar})$ is slightly smaller (but compatible within errors) than the value obtained in Ref.\cite{Sanchis(2022b)}, without tensor modes.

Notice that besides the rather modest improvement of the $\chi^2_{\rm d.o.f.}$ of the fits in Fig.q as new cutoffs are incorporated, the downward tail at large angle is only reproduced in cases (c) and (d), where the odd-parity preference can be traced back to the introduction of the set of infrared cutoffs.
More clearly, figure 2 shows the increasing improvement 
of the fits as new cutoffs are 
successively incorporated to the fit. It is interesting to note that the inclusion of tensor modes especially improves the agreement in the interval 
$10 \lesssim\ \ell\ \lesssim 30$.

\begin{figure*}[h]
\includegraphics[width=13.cm]{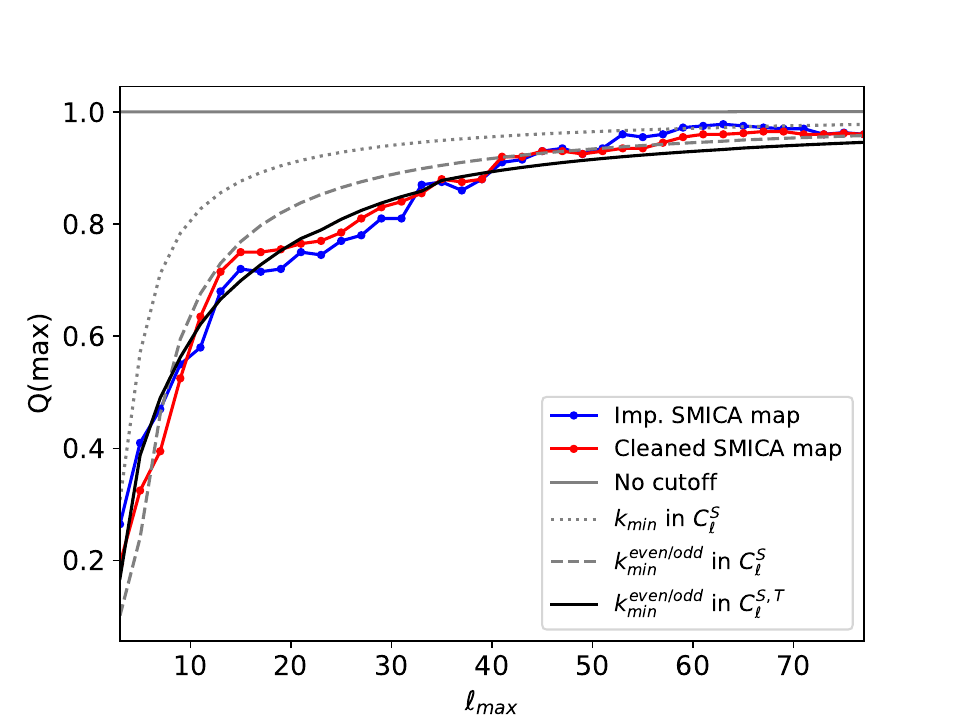}
\caption{\label{fig:3}Parity-statistic $Q(\ell_{max})$ versus $\ell_{max}$as as defined in the text  
under the same
conditions as in caption of Fig.1,  and the
measured points by {\it Planck} (red and blue) under application of two different SMICA masks 
to suppress undesired foreground.}
\end{figure*}

\section{Summary and discussion}

We have examined angular temperature correlations of the CMB, following the trail
of previous works \cite{Melia(2018),Sanchis(2022), Sanchis(2022b)}. We first introduced a couple of 
infrared cutoffs into the scalar power spectrum, thereby
modifying the behaviour of the 2-point correlation function $C(\theta)$ and the parity statistic $Q(\ell_{\rm max})$. In this way we were able to bring the model expectation
closer to the {\it Planck} 2018 data points. However, the
effect on correlations and parity balance is limited to rather low multipoles as $\ell\ \lesssim\ 6$ and the improvement finally achieved is rather modest.

Thus, in order to improve further the fits of both $C(\theta)$ and $Q(\ell_{max})$, tensor modes contributing to the CMB temperature fluctuations 
%(though to a lesser extent than scalar modes) 
were included in the analysis. Let us note that the 
possibility of unraveling the influence of the cosmological
gravitational background on the observed lack of large-angle temperature correlations in the CMB has been envisaged elsewhere, e.g., \cite{Galloni:2023pie}.  

On the other hand, motivated by an 
extra dimension KK model, two sets of infrared cutoffs satisfying $k_{\rm min}^{\rm even} = 2k_{\rm min}^{\rm odd}$ were introduced to both the scalar and tensor
power spectra. The resulting lower cutoffs $u_{\rm min}^{\rm odd/even}$ affect differently
the $C_{\ell\rm odd/even}^{S,T}$ coefficients in the Legendre polynomial
expansion of $C(\theta)$, consequently modifying the fits to angular distributions. The value of $\chi^2_{\rm d.o.f.}$ for the $C(\theta)$ fit goes from 2.5 to 0.95, a substantial improvement with respect to the initial assumption of no cutoff at all. Furthermore, the accordance of the parity statistic $Q(\ell_{max})$ to data largely improves, as can be seen in Fig.2.

Lastly, from our analysis we estimate the value of the tensor-to-scalar ratio $r$ defined in Eq.(\ref{eq:r}). To do so, we employed the ratio 
$C_{\ell}^T/C_{\ell}^S$ for different (low) $\ell$'s computed from our fits of the correlation
function $C(\theta)$. The following expression (from \cite{Mukhanov(2005)}) was used:
\begin{equation}\label{eq:rtheory}
r\ \approx 0.68\ \times\ \frac{C_{\ell}^T}{C_{\ell}^S} 
\end{equation}

Inserting now the average value $\langle C_{\ell}^T/C_{\ell}^S\rangle \simeq 0.04$  computed from our fits in the
$10\leq \ell \leq 20$ interval (where the SW-plateau is reached for both scalar and tensor modes), we get the estimate
\begin{equation}\label{eq:rvalue}
r\ \simeq 0.027\ \pm 0.003
\end{equation}
which lies under current limits~\cite{ParticleDataGroup:2022pth}. Let us point out that the above error bar has been estimated exclusively from the dispersion of the $C_{\ell}^T/C_{\ell}^S$ values obtained from our fits, while other no less important uncertainties, like the theoretical approximations 
and modelling dependence used throughout this paper, 
have not been taken into account. Therefore the above $r$ value  is mainly intended as a consistency test of our results, but
not a precise determination of its value from this study. 
We can also cast the above result into usual slow-roll inflation parameters. For example, using $r=16\epsilon$, we get $\epsilon\ \simeq\ 0.0017$, well within the slow-roll assumption.

As a final remark, uncertainties including the cosmic variance (not considered in this paper), a possible statistical fluke at large angles, contamination or non-cosmological effects~\cite{Creswell(2021),Zhao(2015)}, or even alternative theoretical explanations \cite{Ashtekar(2020)},  
must certainly be kept in mind. Nevertheless, the suggestive accordance between the observed points and fits
achieved when successive sets of infrared cutoffs in the power spectrum (for both scalar and tensor modes) are incorporated into the analysis, is remarkable enough to stress this way of unraveling tensor
modes (from PWG produced during inflation) showing up in CMB
temperature correlations, besides the usual search based on B-mode polarization.

\vspace{1.5cm}
\begin{acknowledgments}
M.A.S.L. thanks interesting discussions with F. Melia and D. G. Figueroa.
 M.A.S.L. acknowledges 
support by the Spanish Agencia Estatal de Investigacion under grant PID2020-113334GB-I00 / AEI
/ 10.13039/501100011033, and by Generalitat Valenciana under grant CIPROM/2022/36. 
The research of V.S. is supported by the Generalitat
Valenciana PROMETEO/2021/083 and the Ministerio de Ciencia e Innovacion PID2020-113644GB-I00. 
\end{acknowledgments}

\appendix*
\section{Even versus odd polynomials and Chebyshev polynomials}

Notice the following relations between Legendre polynomials and the square of cosine functions
with entire and half-entire (2$\pi$) periods playing a fundamental role in the assignments of the infrared cutoffs to even and odd multipole terms respectively:
\begin{eqnarray}\label{eq:Lagrange}
    P_1(\cos{\theta}) & = & -1+2\cos^2{(\theta/2)} \nonumber \\
P_2(\cos{\theta}) & = & -0.5+1.5\cos^2{(\theta)} \nonumber  \\
P_3(\cos{\theta}) & = & -1+0.75\cos^2{(\theta/2)}+1.25\cos^2{(3\theta/2)} \nonumber \\
P_4(\cos{\theta}) & = & -0.7184+0.6249\cos^2{(\theta)}+1.0937\cos^2{(2\theta)} \nonumber \\
P_5(\cos{\theta}) & = & -1+0.4687\cos^2{(\theta/2)}+0.5469\cos^2{(3\theta/2)}+
0.9844\cos^2{(5\theta/2)} \\ 
& \cdots & \nonumber
\end{eqnarray}

Higher order Legendre polynomials replicate the same pattern: even and odd Legendre polynomials   
either contain $\cos^2{[n\theta]}$ or $\cos^2{[(n+1/2)\theta]}$ terms, to be put in correspondence 
with the integer and half-integer modes in the Fourier
decomposition of the
fluctuating field under cyclic conditions. Consequently, the infrared cutoffs 
$k_{\min}^{\rm odd/even}$ apply to even and odd multipole
terms respectively, in the expansion of the correlation function $C(\theta)$ in terms of Legendre polynomials. 
For more details see \cite{Sanchis(2022b)}.

Alternatively, one can write the above relations in terms of Chebyshev polynomials $T_n(\cos{(\theta)})= \cos{(n\theta)}$
~\cite{Abramowitz(1970)}),
\begin{eqnarray}\label{eq:Chebyshev}
    P_1(\cos{\theta}) & = & T_1 \nonumber \\
P_2(\cos{\theta}) & = & 0.25+0.75T_2 \nonumber  \\
P_3(\cos{\theta}) & = & 0.375T_1+0.625T_3 \nonumber \\
P_4(\cos{\theta}) & = & 0.1409+0.3124T_2+0.5468T_4 \\
& \cdots & \nonumber
\end{eqnarray}
Again it becomes apparent the relationship between the (even or odd) parity of the Legendre polynomials and the respective infrared cutoffs used to compute the multipole coefficients $C_{\ell\ \rm even/odd}$  through the
even and odd modes of the Fourier expansion of fields.

\nocite{*}
%\bibliography{unsrt}% Produces the bibliography via BibTeX.

\end{document}